\documentclass[aps,prb,twocolumn,nofootinbib]{revtex4}
\usepackage{epsfig}
\usepackage[dvipsnames,usenames]{color}
\usepackage{enumerate}
\usepackage{color}
\usepackage{colordvi}
\usepackage{amsmath}
\usepackage{amssymb}

\emergencystretch=\maxdimen
\begin{document}

\title{Finite $f$-Electron Bandwidth in a Heavy Fermion Model}

\author{A. Euverte$^1$,
S.  Chiesa$^2$, R.T.~Scalettar$^3$, and
G.G. Batrouni$^{1,4,5}$}

\affiliation{$^1$INLN, Universit\'e de Nice-Sophia Antipolis, CNRS;
1361 route des Lucioles, 06560 Valbonne, France }
  
\affiliation{$^2$ Department of Physics, College of William \& Mary,
Williamsburg, VA 23185, USA}

\affiliation{$^3$Physics Department, University of California, Davis,
California 95616, USA}

\affiliation{$^4$Institut Universitaire de France, 103 bd Saint-Michel,
75005 Paris, France,}

\affiliation{$^5$Centre for Quantum Technologies, National
University of Singapore; 2 Science Drive 3 Singapore 117542}

\begin{abstract}
  Determinant Quantum Monte Carlo (DQMC) is used to study the effect
  of non-zero hopping $t_{f}$ in the ``localized" $f$-band of the
  periodic Anderson model (PAM) in two dimensions.  The low
  temperature properties are determined in the plane of interband
  hybridization $V$ and $t_{f}$ at fixed $U_{f}$ and half-filling,
  including the case when the sign of $t_{f}$ is opposite to that of
  the conduction band $t_{d}$.  For $t_f$ and $t_d$ of the same sign,
  and when $t_f/ t_d> (V/4t_d)^2$, the non-interacting system is
  metallic.  We show that a remnant of the band insulator to metal
  line at $U_f=0$ persists in the interacting system, manifesting
  itself as a maximal tendency toward antiferromagnetic correlations
  at low temperature.  In this ``optimal'' $t_f$ region, short range
  (e.g.~near-neighbor) and long-range spin correlations develop at
  similar temperatures and have comparable magnitude.  Both
  observations are in stark contrast with the situation in the widely
  studied PAM ($t_f=0$) and single band Hubbard model, where short range 
  correlations are stronger and develop at higher temperature.  The effect
  that finite $t_f$ has on Kondo screening is investigated by
  considering the evolution of the local density of states for
  selected $t_f$ as a function of $V$. We use mean field theory as a
  tool to discriminate those aspects of the physics that are genuinely
  many-body in character.
\end{abstract}

\maketitle

\section{Introduction}

The single band Hubbard Hamiltonian is the simplest itinerant electron
model used to describe the effects of strong correlation in solids
\cite{rasetti91,montorsi92,gebhard97,fazekas99}.  At half-filling and
at low temperatures, an on-site repulsion $U$ drives the emergence of
a Mott insulator (MI) phase as well as, on bipartite lattices,
antiferromagnetic (AF) correlations.  Upon doping, mobile defects are
introduced into this MI, and the Hubbard model exhibits more exotic
correlation effects, including, in two dimensions, incommensurate
charge and spin correlations (``striped phases")
\cite{zaanen89,inui91,yang91,machida89,kato90} and, very possibly, $d$-wave
superconductivity \cite{scalapino94}.

Some fundamental correlation physics, however, is best considered
within multi-band Hamiltonians.  The Periodic Anderson model (PAM)
\cite{fazekas99}, for example, describes the competition of
antiferromagnetic order and singlet formation in which highly
correlated electrons in one ``$\,f\,$" orbital are screened by weakly
correlated electrons in a second ``$\,d\,$" orbital.  These two
possible ground states in the PAM are thought to describe
qualitatively the observation that certain $f$-electron systems like
CeAl$_2$ are antiferromagnetic, while others like CeAl$_3$ are not,
and also to be relevant to materials in which AF is lost (and
superconductivity appears) as pressure is applied and the ratio of
correlation to kinetic energies is decreased\cite{sugitani06}.  The
``heavy fermion" \cite{stewart01} behavior in such rare earth
materials, in which the electrons acquire a large effective mass, is
also a feature of the PAM.  Some physics which was believed to be
specific to the PAM, notably the appearance of a Kondo resonance in
the density of states, is now known to occur in single band models as
well \cite{georges96,vollhardt93,ulmke96,held00}.

Indeed, analytic and numerical investigations have shown that the
AF-singlet competition is rather generic.  That is, it occurs in a
variety of other two band models, not just ones in which the $f$-band
is completely localized, $t_{f}=0$.  Perhaps the most straightforward
alternative to the two dimensional PAM is the two band (or two layer)
Hubbard Hamiltonian,	 in which the intralayer hopping and
interaction strengths are chosen to be identical (i.e.~$t_f=t_d$ and
$U_f=U_d$).  Although there is initially a greater AF tendency than in
a single layer as the inter-layer hybridization, $V$, grows toward
$V=t_f$, this is followed by a rather rapid decrease of magnetization
and the eventual loss of long range order when $V\simeq 1.6
t_f$\cite{scalettar94}.  This AF to paramagnetic (singlet) transition
is also present in quantum spin models like the two layer Heisenberg
model \cite{sandvik94}, and in Hubbard and Heisenberg ladders
\cite{white92-98}.

The goal of the present paper is to obtain a more systematic picture
of the nature of the AF-singlet competition and related aspects of
Kondo physics at finite temperature.  Specifically, we consider a two
orbital model in which $t_{f}$ interpolates smoothly between the
well-known PAM ($t_{f}=0)$ and the equal bandwidth ($t_{f}=t_{d}$)
cases.  We also study $t_{f}<0$, a regime in which, as at $t_{f}=0$,
the system is a band insulator in the absence of interactions.  A
particularly interesting issue is the interplay of the RKKY
interactions, in which the $f$ moments couple indirectly through the
conduction bands, and direct exchange $J \sim t_{f}^2/U_{f}$.  Both
give rise to antiferromagnetic correlations, yet we show that their
joint effects do not manifest as a straightforward reinforcement.  One
origin of the complexity is that while $V$ increases the RKKY AF
tendency, it also affects the band structure in the localized $f$-band
and hence the density of states there.  This latter effect is captured
by the Stoner criterion in a mean field treatment.

Allowing a finite bandwidth in the $f$-band makes the PAM a more
realistic model for describing heavy-fermions materials. In the case
of actinides \cite{PhysRevB.76.155126}, the ratio of the inter-band hybridization over 
the $f$-band nearest-neighbor hopping
has been recently estimated to be of order $V/t_f\approx3$.
Several papers have considered similar models
\cite{PhysRevB.49.4432,JPSJ.74.2517,JPSJ.80.064710,JPSJ.69.1777,demedici05},
but the effect of $t_f$ has been systematically explored only in the
infinite dimension case \cite{JPSJ.69.1777} using dynamical mean field
theory (DMFT).  For example, de' Medici {\it et al.} \cite{demedici05}
have focused on the closing of the Mott gap with increasing $t_f$ and
how this insulator-metal transition differs at zero and finite
temperature.  We will show that accounting for magnetism leads to
different conclusions from the mere renormalization of the
non-interacting density of states found by DMFT \cite{JPSJ.69.1777}.

Although our focus is on the magnetic correlations and
spectral function in the $f$-band, our results connect also to work
which explores the question of `orbitally selective' Mott transitions
\cite{liebsch04,liebsch05,arita05,inaba06,costi07,koga04,
  ferrero05,demedici05b,ruegg05,inaba05,knecht05,biermann05}.  Here
two band Hamiltonians with different $U_f$ and $U_d$, and which are
coupled either by inter-orbital hybridization (as in the present
manuscript) or interactions, are solved.  The key issue is whether the
two fermionic species can be of mixed character, with one metallic and
one insulating.

The remainder of this paper is organized as follows: In Section II, we
write down the model Hamiltonian to be examined, and briefly summarize
the mean field theory (MFT) and DQMC formalisms.  Section III presents
MFT and DQMC results for magnetic correlations, while Sec.~IV those
for the local spectral function in the correlated band.  The summary
and conclusions are in Section V.

\section{Model and Calculational Methods}

The two band fermionic Hubbard Hamiltonian we consider here,

\begin{eqnarray}
  \hat {\mathcal H} = &-&\sum_{\langle {\bf j},{\bf k} \rangle l \sigma } 
  t^{\phantom{\dagger}}_{l} 
  (c_{{\bf j} l\sigma}^\dagger c_{{\bf k}l\sigma}^{\phantom{\dagger}} + {\rm h.c.}) \\
  &-& V \sum_{{\bf j},\sigma}(c_{{\bf j}f\sigma}^\dagger c_{{\bf
      j}d\sigma}^{\phantom{\dagger}} 
  + {\rm h.c.}) \nonumber \\
  &+& \sum_{{\bf j}l}U^{\phantom \dagger}_{l}(n_{{\bf j}l}^{\uparrow}-\frac{1}{2})
  (n_{{\bf j}l}^{\downarrow}- \frac{1}{2})
  -\mu \sum_{{\bf j}l\sigma} n_{{\bf j}l}^{\sigma} 
  \,\,\, ,
  \nonumber
\label{eq:hamiltonian}
\end{eqnarray}
describes a two-dimensional square lattice with electronic bands $l=d$
and $l=f$. The coordinates $({\bf j}\,l)$ label the spatial site and
band respectively; $\sigma \in \{\uparrow,\downarrow\}$ denotes the
spin of the electron. The operators $c^{\phantom{\dagger}}_{{\bf
    j}l\sigma}$, $c^{\dagger}_{{\bf j}l\sigma}$ and $n_{{\bf
    j}l}^{\sigma}=c^{\dagger}_{{\bf
    j}l\sigma}c^{\phantom{\dagger}}_{{\bf j}l\sigma}$ are the
destruction, creation and number operators. The first terms are the
intra-band and inter-band kinetic energies.  The nearest-neighbor
hopping matrix element in the $d$-band will be chosen to be the unit
of energy in the remainder of this work, $t_{d}=1$, while its $f$-band
counterpart, $t_{f}$, will be allowed to vary in the range
$[-1.2,1.2]$. Because properties at $V$ and $-V$ are related by the
transformation $c_{id\sigma}\rightarrow -c_{id\sigma}$, we restrict
the range of studied $V$ to only positive values.  The on-site
repulsion $U_l$ is chosen to be constant within a band: the $f$-band
will include a moderate interaction $U_f\equiv U=4$ (unless otherwise
mentioned), while the $d$-band will be non-interacting $U_d=0$.  The
chemical potential $\mu$ is set to zero, the system being then at
commensurate filling for both $l=d,f$.  This choice emphasizes Mott
and antiferromagnetic physics, and also avoids any sign problem in our
DQMC simulations.

\begin{figure}[ht]
\centerline{\includegraphics[width=8.5cm,angle=0]{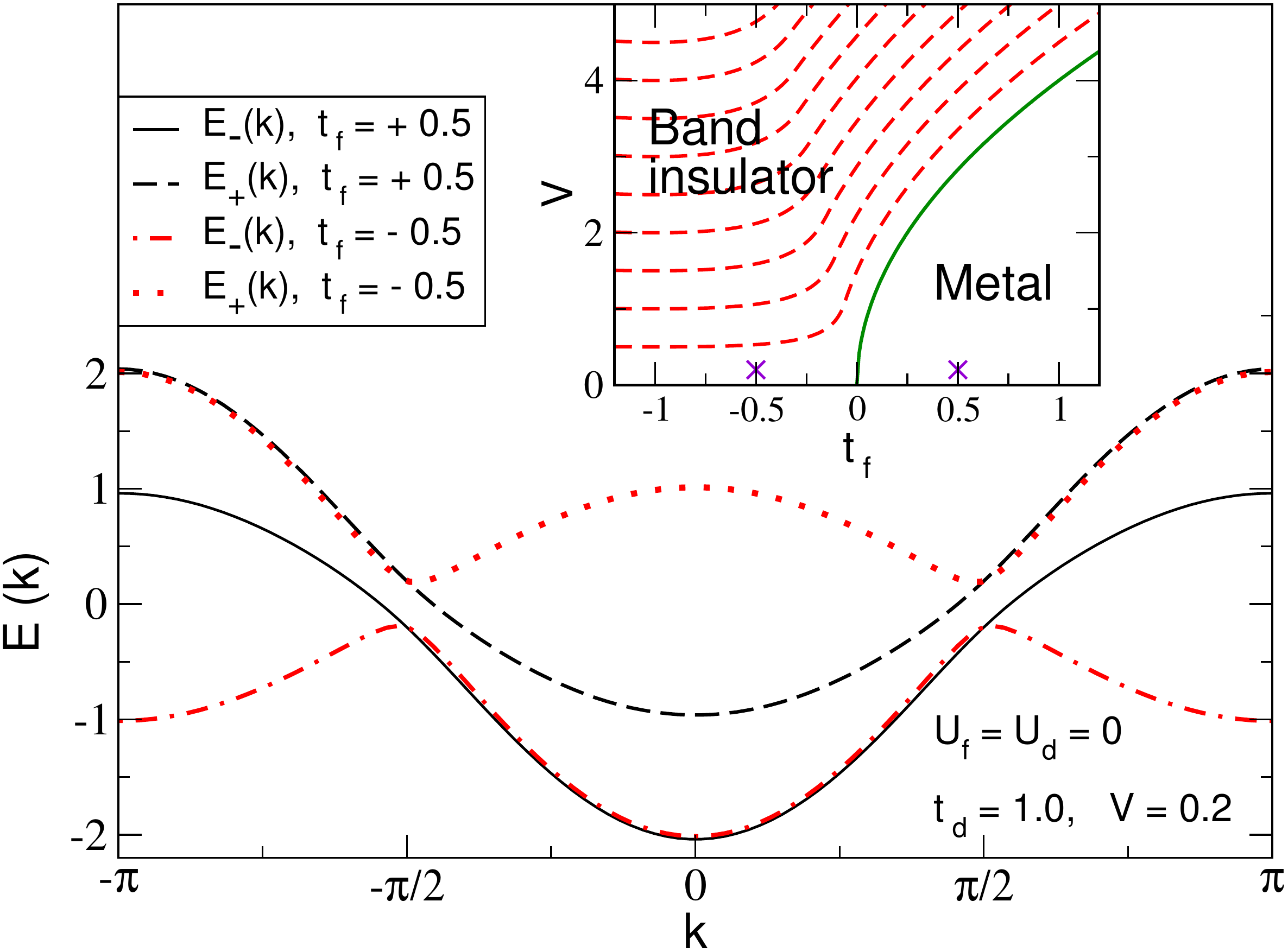}}
\caption{The $U_l=0$ band structure (see Eq.~\ref{eq:bs}),
  illustrating the distinct behavior of the two cases when $t_d$ and
  $t_f$ have the same (opposite) sign.  In the former situation, there
  are two overlapping bands as long as the interband hopping $V$ is
  not too large, and the half-filled system is a metal.  In the latter
  case, a gap opens between the bands and the system is an insulator
  at half-filling, for all $V$. The inset shows the resulting non-interacting
  phase diagram, in which the dashed lines are isolines at integer
  values of the gap.}
\label{fig:opphop_pedagogy}
\end{figure}

In the non-interacting limit $U=0$, the regions with positive and
negative $t_{f}$ have quite distinct behaviors. The former is a metal
with two overlapping bands, up to a critical inter-band hybridization
above which it crosses into a band insulator, while the latter is
always a band insulator. Such behavior is easily understood in terms
of the dispersion and ensuing crossing of the independent $f$- and $d$-
bands, and is illustrated in the main panel of
Fig.~\ref{fig:opphop_pedagogy} for a one dimensional geometry.  More
specifically, setting $x(k)=\cos k_x+ \cos k_y$ gives band energies of
the form
\begin{eqnarray}
E^0_{\pm}(x)=-(1+t_f)x\pm\sqrt{(1-t_f)^2x^2+V^2}, 
\label{eq:bs}
\end{eqnarray}
with a metal-insulator transition at $|V|=4\sqrt{t_f}$.  Because there
is perfect $(\pi,\pi)$ nesting between the two branches of the Fermi
surface, the metal is expected to develop long-range magnetic order at
$T=0$ and for arbitarily weak repulsion.  On the other hand,	 the
presence of a gap in the band insulator implies that a finite $U$ is required for the
system to develop AF order. The gap is given by $\Delta=2 E^0_+(2)$
when $t_f>0$ or when $V(1+t_f)>4\sqrt{-t_f}(1-t_f)$ and $t_f<0$, and
by $\Delta=4V\sqrt{-t_f}/|1-t_f|$ in all remaining cases of interest
in the paper.

\subsection{Mean field theory}
We use the conventional decoupling of the on-site interaction $U
n^\uparrow n^\downarrow \rightarrow U ( \, n_\uparrow \langle
n_\downarrow \rangle + \langle n_\uparrow \rangle n_\downarrow -
\langle n_\uparrow \rangle \langle n_\downarrow \rangle \,)$ with
ansatz
\begin{equation}
\begin{split}
  \langle n_{{\bf j}l\uparrow} \rangle + \langle n_{{\bf j}l\downarrow} \rangle &= 1 \\
  \langle n_{{\bf j}l\uparrow}\rangle - \langle  n_{{\bf j}l\downarrow} \rangle &= 2 m_l \\
\end{split}.
\end{equation}
This suffices since, at half-filling, we do not expect the occurence
of non-collinear or inhomogeneous phases.  By introducing the
mean-field eigenvalues of a single layer
\begin{equation}
  E_{k\pm}=\pm\sqrt{\epsilon_k^2+(Um)^2},
\end{equation}
we can conveniently express those of the bilayer as
\begin{eqnarray}
  \lambda &=& \pm\frac{1}{2}\Big[E_{k,d}^2 + E_{k,f}^2+2V^2  \nonumber \\
  && \hspace{.8cm} \pm \sqrt{(E_{k,d}^2-E_{k,f}^2)^2+4V^2\mathcal{E}^2}\Big]^{1/2} 
\label{eq:lambda}
\end{eqnarray}
where we omitted the $\pm$ subscript (since $E_{k\pm}$ always enter as
squares) and we defined
\begin{equation}
  \mathcal{E}^2=\left(\sum_l \epsilon_{k,l}\right)^2+\left(\sum_l m_l U_l\right)^2.
\label{eq:epsilonsq}
\end{equation}
It is easy to verify that when $m_f m_d<0$ (interband
anti-ferromagnetic order) the equation $\lambda=0$ does not admit any
real solution regardless of the value of $k$.  This implies that, when
order sets in, the Fermi surface is fully gapped.

\subsection{Determinant Quantum Monte Carlo}
In this approach\cite{blankenbecler81,white89} the partition function
is written as a path integral and the interaction is decoupled through
the introduction of a space and imaginary time dependent auxiliary
field.  Sampling this field stochastically produces the exact physics
of the underlying Hamiltonian on finite clusters, apart from
statistical errors which can be reduced by running the simulation
longer, and ``Trotter errors"\cite{trotter} associated with the
discretization of the inverse temperature $\beta$.  These can be
eliminated by extrapolation to zero imaginary time mesh size $\delta
\tau$.  Here we have set $\delta \tau=1/(8 t_{d})$ and verified that
our results are qualitatively unchanged when $\delta \tau$ is reduced.

The DQMC results we present are computed for lattices of $N=8\times8$
sites, and two bands.  At various points of the phase diagram, we
checked that larger clusters do not lead to qualitatively different
results.  In many cases our focus is on physics at short length
scales, e.g.~near-neighbor spin correlations, which converge rapidly
with lattice size.  Every data point was obtained by averaging several
independent simulations performed over a set of four different
boundary conditions, leading to a better sampling of the first
Brillouin zone, and thereby reducing finite size effects.

Magnetism is measured by examining the spin correlation in the
$f$-band, $\langle {\vec{\sigma}}_{{\bf j}\,f } \cdot
\vec{\sigma}_{{\bf j}+{\bf r},f} \rangle$ with $\vec{\sigma}$ given by
\begin{eqnarray}
  \sigma^z_{{\bf j}} &=& c_{{\bf j}\uparrow}^\dagger c_{{\bf
      j}\uparrow}^{\phantom{\dagger}} -c_{{\bf j}\downarrow}^\dagger
  c_{{\bf j}\downarrow}^{\phantom{\dagger}}; \phantom{aa} \sigma^+_{{\bf
      j}} = c_{{\bf j}\uparrow}^\dagger c_{{\bf
      j}\downarrow}^{\phantom{\dagger}}; \phantom{aa} \sigma^-_{{\bf j}}
  = c_{{\bf j}\downarrow}^\dagger c_{{\bf
      j}\uparrow}^{\phantom{\dagger}} \nonumber.
\end{eqnarray}
We focus on short range (near neighbor) correlations $\langle
{\vec{\sigma}}_{{\bf j}\,f } \cdot \vec{\sigma}_{{\bf j}+\hat x,f}
\rangle$ and long range order probed via the antiferromagnetic
structure factor $S_f^{\rm af}$ defined in Eq.~\ref{eq:saf}.

\begin{figure}
\centerline{\includegraphics[width=8.5cm,angle=0]{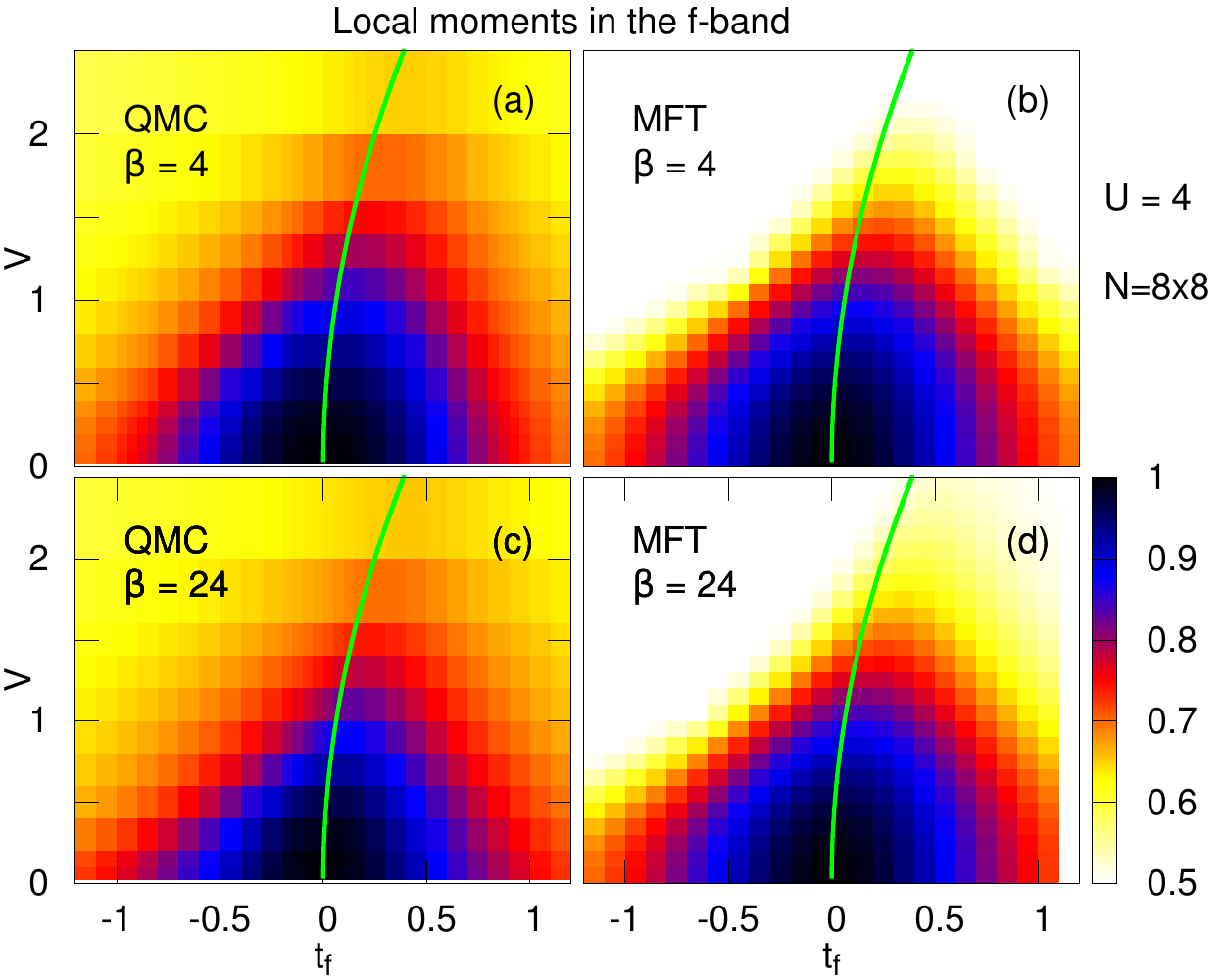}}
\caption{Evolution of the local $f$ moment $m_f$ with $t_f$ and $V$,
  at $U=4$, obtained by QMC (left panels) and MFT (right panels).  At
  $\beta =4$ (top panels), the moments are mostly formed and
  decreasing the temperature to $\beta=24$ has no significant effect.
  $m_f$ is largest for small values of these two terms, since this
  minimizes the quantum fluctuations.  The curve extending upwards
  from the origin $(t_f,V)=(0,0)$ is the noninteracting ($U=0$)
  band-insulator to metal transition line.  In the $U$ nonzero case,
  $m_f$ is best formed along this line.  }
\label{fig:m0} 
\end{figure}

\section{Magnetic Correlations}
The on-site repulsion $U$ favors localization of electrons by impeding
double occupancy. We study this tendency by showing the local moments
$m_f$ in the $f$-band in Fig.~\ref{fig:m0}.  $m_f$ is related to the
double occupancy $D_f$ by:
\begin{eqnarray*}
  m_f =\frac{1}{N}\sum_{\bf j} \langle (\sigma^z_{{\bf j}{f}})^2\rangle 
  = 1 - \frac{2}{N}\sum_{\bf j} \langle n_{{\bf j}{f}}^{\uparrow} n_{{\bf j}{f}}^{\downarrow}\rangle = 1-2D_f	.
\end{eqnarray*}
The most obvious general trend is that localization decreases upon
increasing the magnitude of either the $f$-band hopping $t_f$ or the
interband hybridization $V$.  This observation of course reflects the
competition between kinetic energy scales and on-site repulsion $U$.
As expected, the MFT local moment vanishes much more abruptly than its
QMC counterpart, reflecting the fact that the moment can act as an
order parameter within MFT.

We can understand the shape of the local moment dome from a weak
coupling perspective in three simple steps. 1) At $T = 0$ and for very
weak interaction strengths, the diverging susceptibility implies that
order must exist in the part of the $t_f-V$ plane where the
non-interacting system is metallic:
the dome would coincide precisely with the region to the right of the green line in
the figures. 2) Finite $U$ values at $T = 0$ cause the insulating
phase to gradually order. In particular, the gap $\Delta$ decreases as $V$ is
decreased at constant $t_f$ (see isolines in Fig. 1)and a transition to an ordered phase
happens when $\Delta\propto U$ {\em i.e.} the dome described in 1)
acquires a “tail” in the negative $t_f$ region. The larger $U$ is, the
thicker the tail becomes. 3) At a finite $T$, this picture needs to be
modified to take into account that order will persist in a given
region only up to a temperature of the order of the $T = 0$ AF
gap. Because the gap is the smallest in the region of large and
positive $V$ and $t_f$, such region is also the first to lose order
as the temperature is raised. These three arguments rationalize the shape of
the moment dome in the $t_f-V$ plane, its asymmetry with respect to
the $t_f = 0$ axis and its apex at positive $t_f$.  In particular,
this asymmetry in the values of $m_f$ implies that a small positive
hopping $t_{f}$ tends to strengthen the moment while a small negative
one tends to weaken it.

\begin{figure}
\centerline{\includegraphics[width=8.5cm,angle=0]{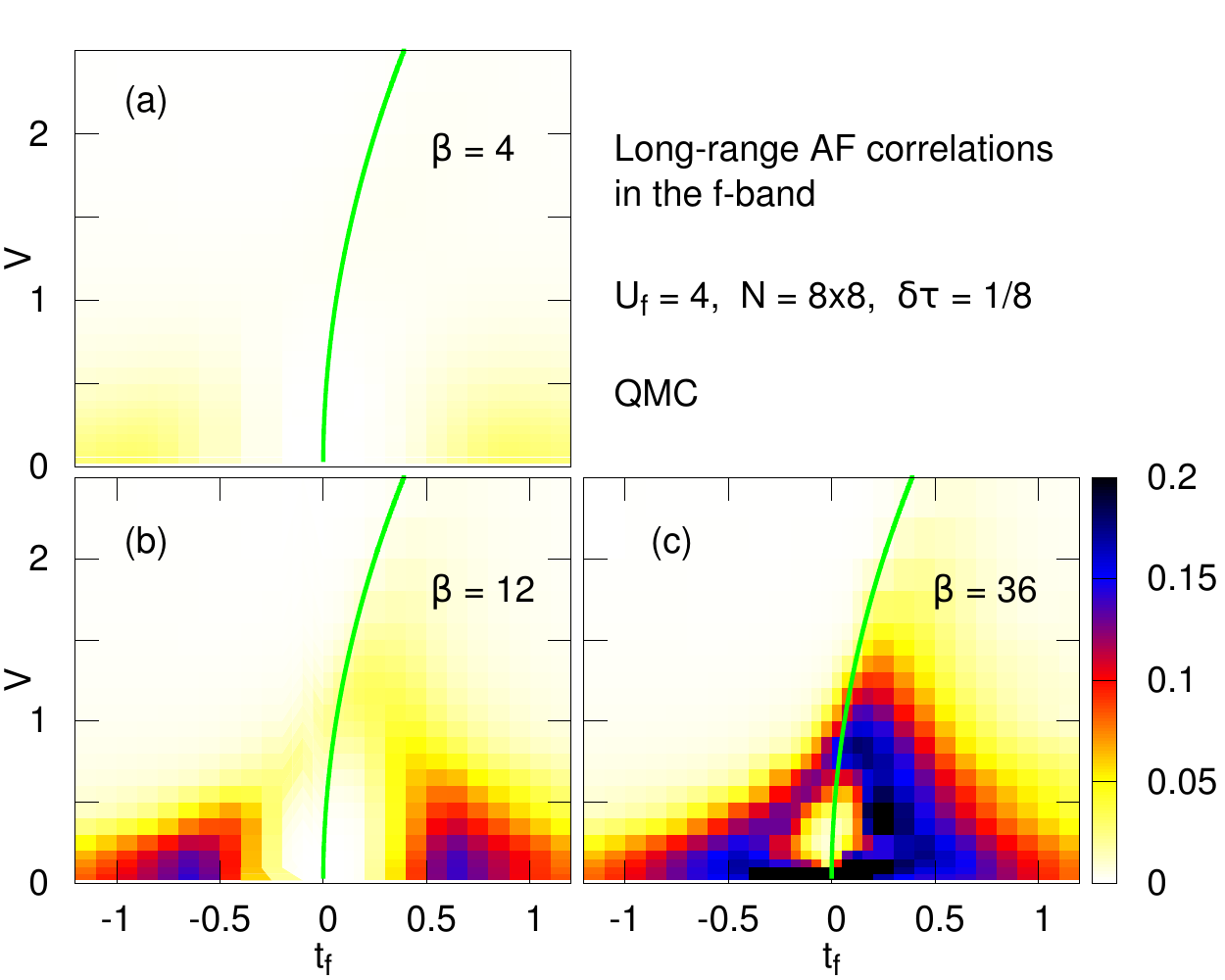}}
\caption{The $f$-band long-range antiferromagnetic correlations shown
  in the $t_f$-$V$ plane for different inverse temperatures $\beta$.
  Because $S^{\rm af}_f$ probes spin ordering at large distance, the
  convergence with increasing $\beta$ is more gradual than for the
  local moments.  As $\beta$ increases, in the paramagnetic region
  where $V$ and $t_{f}$ are both small so too are the AF energy
  scales, and AF order is absent.  This region shrinks as $\beta$
  increases and the temperature systematically falls below the AF
  energy scales.  As with the local moment, $S^{\rm af}_f$ is peaked at
  small positive $t_{f}$.  The green curve is the noninteracting
  metal-insulator transition line.  }
\label{fig:Saf0} 
\end{figure}

In regimes where a weak coupling treatment is appropriate {\em
  i.e.}~when moments are small, these arguments carry through to the
staggered magnetization and, as a direct consequence, to long range
correlations (although, because the system is two-dimensional, the
former is only different from zero at $T=0$).  We quantify the
evolution of long-range correlations by looking at
\begin{eqnarray}
  S^{\rm af}_{f} &=& \frac{1}{3N'} \sum_{{\bf j,k}} ^{\prime} \, \langle \,
  \sigma^z_{{\bf k}\,f} \sigma^z_{{\bf j}\,f} +
  2\,\sigma^-_{{\bf k}\,f}
  \sigma^+_{{\bf j}\,f} \, \rangle \, (-1)^{|{\bf k - j}|}.
\label{eq:saf}
\end{eqnarray}
This quantity is related to the antiferromagnetic structure factor,
but the prime symbols in the sum and in the number of sites $N'$
indicate that we omitted contributions from local and nearest neighbor
correlations in order to single out better the long range behavior
\cite{varney09}.  As shown in Fig.~\ref{fig:Saf0}, $S^{\rm af}_f$
defines a dome with the same characteristic asymmetry as that of the
moments, but whose edge is more sharply defined, reminiscent of the
behavior of an order parameter.

\begin{figure}
  \centerline{\includegraphics[width=8.5cm,angle=0]{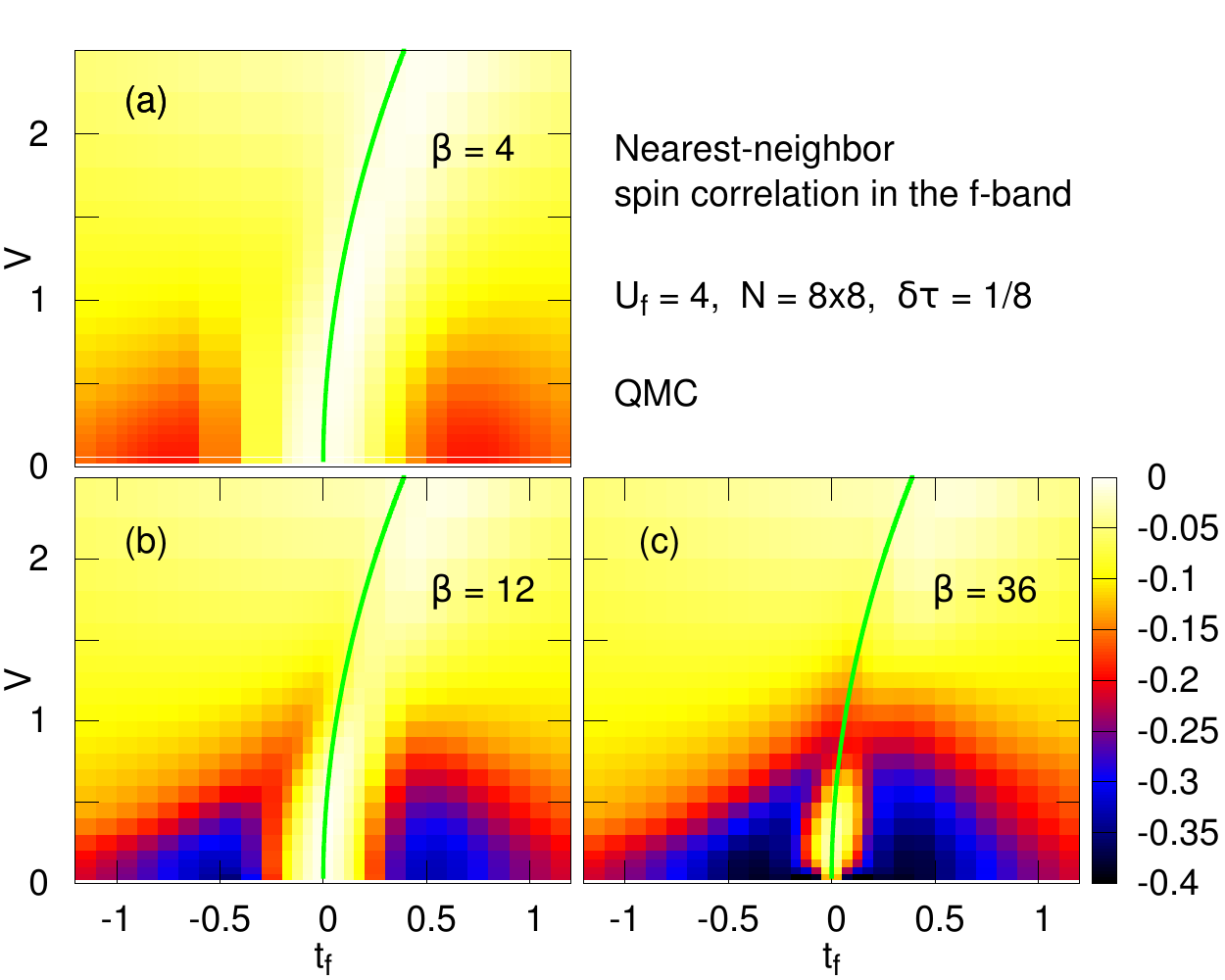}}
  \caption{ Near neighbor spin correlations in the $f$-band.  $t_f$
    provides an exchange interaction $J=4t_f^2/U$ in the $f$-band,
    which leads to antiferromagnetic correlations.  $V$ also causes
    antiferromagnetic interactions via RKKY coupling.  There is
    significant growth in the spin correlations as $\beta$ increases,
    even well after the local moment has saturated
    (Fig.~\ref{fig:m0}).  The $U=0$ band insulator to metal transition
    line is characterized by reduced values of the near neighbor spin
    correlations.  }
\label{fig:SX01} 
\end{figure}

At strong coupling, where $U$ is large and both $t_f$ and $V$ are
small, the weak coupling picture needs to be modified in favor of one
where the local moments are fully formed and interact with conducting
electrons and each other via, respectively, exchange couplings
$J_\perp\propto V^2/U$ and $J\propto t_f^2/U$.  To develop long range
correlations in these regimes one needs to get down to temperatures of
the order of $J$ and $J_\perp$ and this, in turn, leads to the
persistence of an area inside the dome where long range correlations
have still not developed at the lowest $T$ we considered.

Having described the behavior of the local moment (Fig.~\ref{fig:m0})
and antiferromagnetic structure factor (Fig.~\ref{fig:Saf0}) we now
turn to the near neighbor spin correlations in the $f$-band, $\langle
{\vec{\sigma}}_{{\bf j}\,f } \cdot \vec{\sigma}_{{\bf j}+1,f} \rangle$
(Fig. ~\ref{fig:SX01}).  One naturally expects the dome of near
neighbor (n.n.) spin correlations to resemble the one of longer range
correlations {\em i.e.} for a given temperature, regions with large
n.n. spin correlations correspond to regions with large long-range
correlations. One would also expect near neighbor correlations to be
always significantly larger than longer range correlations, rather
independently of temperature, and, in fact, much larger at high $T$
where long range correlations are exponentially small.  Although these
expectations are satisfied in much of the $t_f-V$ plane and at low
$T$, our results indicate that the finite $T$ scenario as the metallic
phase is entered is of less straightforward interpretation.

\begin{figure}
\centerline{\includegraphics[width=8.5cm,angle=0]{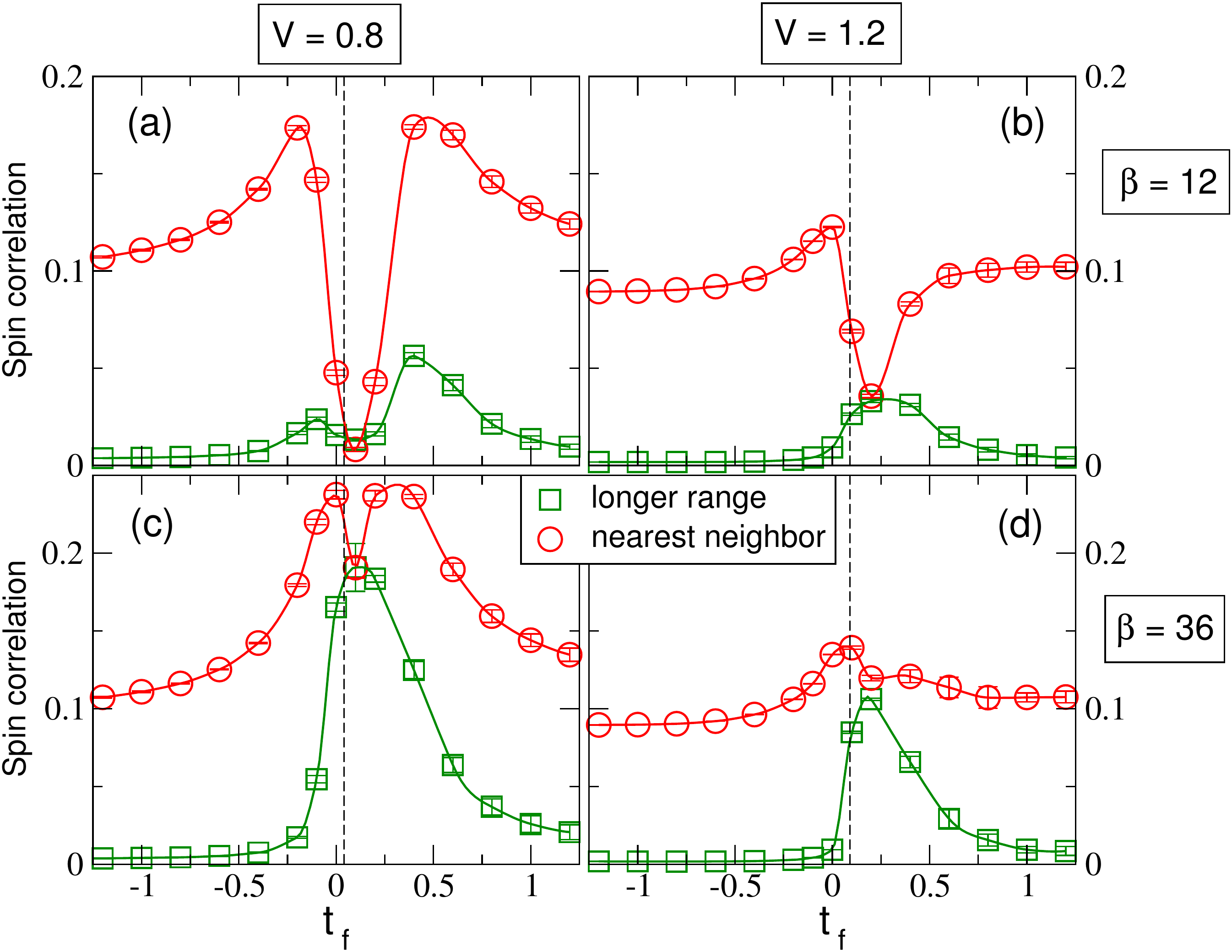}}
\caption{ Comparison of the nearest-neighbor spin correlation in the
  $f$-band against the long-range spin correlations defined in
  Eq.~\eqref{eq:saf}.  Results are shown as functions of $t_f$ for two
  values of the hybridization $V=0.8,1.2$, and at two distinct
  temperatures $\beta=12,36$. The dashed lines indicate the
  metal-insulator transition in the non-interacting case, and closely
  track the downturn of the magnitude of n.n. spin correlation.  }
\label{fig:short_long} 
\end{figure}

We can more precisely illustrate the anomalous behavior of the n.n.
correlations along the metal-insulator line by looking at the
evolution of short and long range correlations for constant $V$ as $t_f$ is
varied (Fig. \ref{fig:short_long}).  For instance, at $V=0.8$ and
$\beta=12$, both correlations show a minimum after
the metal-insulator line is crossed. 
As the temperature is decreased to $\beta=36$ and in proximity to the same value of $t_f$,
longer range correlations have developed a peak while the n.n.~ones still 
show a dip.  At this low temperature, both correlations are of essentially equal magnitude.
This is the rather generic behavior found in
correspondence to crossing the metallic line (see $V=1.2$ in the figure), 
which contradicts both
expectations above.

It is impossible to attribute this effect to the
$f$-intralayer exchange coupling because the latter is too small and the
temperature too high.  Instead, the correlation between the position
of the peak and the metal-insulator line suggests that the change brought by 
$t_f$ must be related to the fact that the underlying non-interacting system
develops a Fermi surface. 
Although this is beneficial to both RKKY interaction
and Kondo screening, it is hard to reconcile our results with a
scenario where the Kondo effect is important since screening of
the local moments should cause a rather uniform decrease of spin 
correlations irrespectively of the distance. 

Therefore, in order to rationalize
the drop in n.n correlations, one must first conclude that the spin-spin interaction 
at the metal-insulator line is weaker than in neighboring regions of the 
$t_f-V$ plane.  At the same time though, to explain
the peak in longer range correlations and the fact that their magnitude
is identical to the n.n. one, one must also conclude that the range
of the effective spin-spin interaction is longer in proximity of the
metal-insulator line than at any other value of $t_f$. 

\section{Density of States}

\begin{figure*}
\centerline{\includegraphics[width=18cm,angle=0]{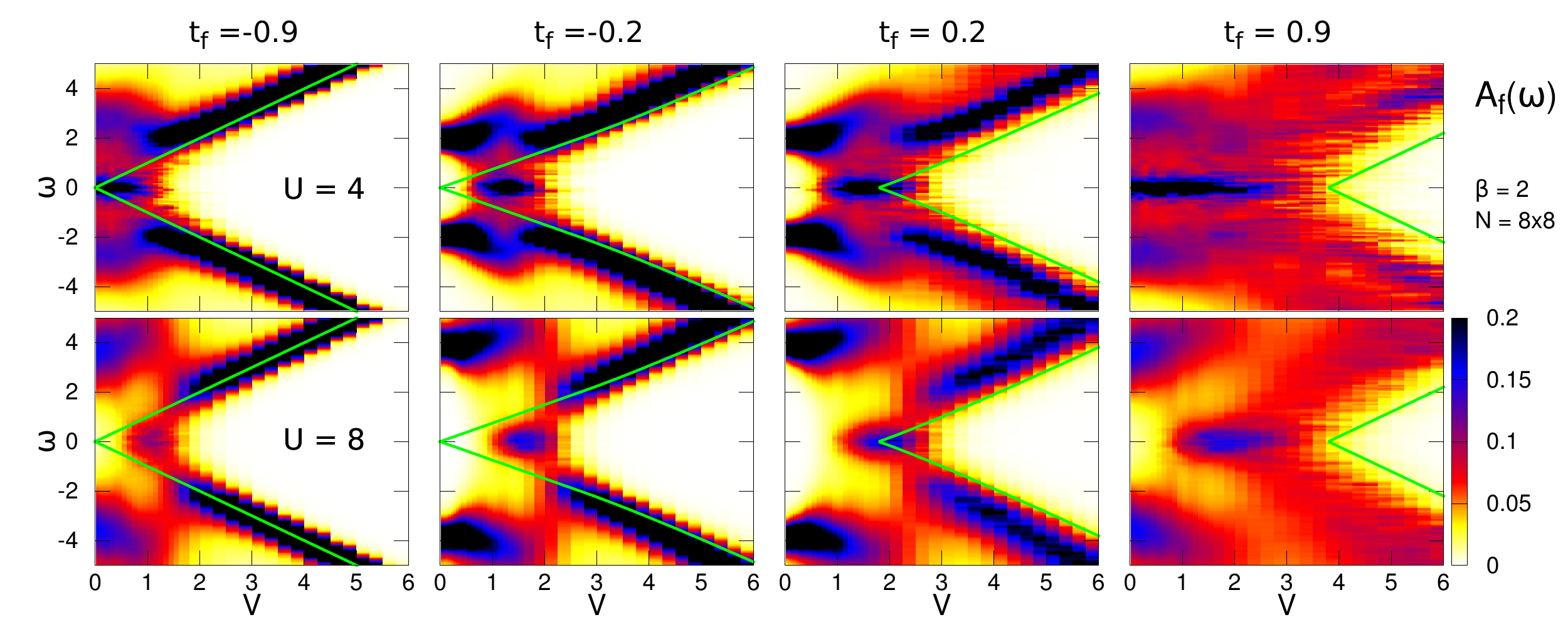}}
\caption{ The spectrum of the $f$-band at $U=4,8$ (top and bottom
  respectively) as function of interband hybridization $V$ for inverse
  temperature $\beta=2$.  Four values of $f$-band hopping $t_f$ are
  shown.  At small $V$, two bands are separated by a Slater gap of
  width $U$ corresponding to the cost of double occupancy.  The Kondo
  resonance is visible at $\omega=0$ for $V$ below a critical
  threshold $V_c$.  The triangular shaped regions at large $V$
  correspond to the singlet formation that prevents Kondo resonance
  and leads to a gap that is well described by the non-interacting
  dispersion relation.  The green lines correspond to the edges of the
  non-interacting density of states, which delimit the band gap.  }
\label{fig:Awbeta2} 
\end{figure*}

The spectrum of excitations of the PAM at half-filling displays
complex features reflecting the interplay of antiferromagnetic order,
the formation of a Mott insulator, and the emergence of Kondo
singlets.  The former two effects suppress the density of states at
the Fermi level, giving rise to a `Slater' or `Mott' gap respectively.
The latter causes screening of the local moments in the $f$-band and
is associated with a Kondo resonance (peak) at the Fermi level.  The
temperature affects these competing possibilities.  For example, a
Kondo resonance might first form as $T$ is lowered, followed by a
splitting of that resonance as magnetic correlations grow.  In
addition to these correlation effects, the spectral function is also
influenced by the character of the non-interacting band structure and,
in particular, by whether or not the system is metallic.

We explore these issues by extracting the single-particle excitation
spectrum $A_f(\omega)$ via analytic continuation of the local
time-dependent Green function $G_f(\tau)=\langle
c^{\phantom{\dagger}}_i(\tau) c^{\dagger}_i(0)\rangle$ measured in QMC
simulations.  This involves the inversion of the relation
\begin{eqnarray} G_f(\tau) = \int_{-\infty}^{+\infty} d \, \omega \,
\frac{e^{-\omega \tau}}{1+e^{-\beta \omega}} \, A_f(\omega),
\end{eqnarray} which we perform using the maximum entropy method
\cite{gubernatis91,beach04}.

We begin with the spectra at inverse temperature $\beta=2$, which is
cold enough to have allowed for some of the many body physics to have
occured, {\it e.g.} moment formation and moment screening, but not
sufficiently cold for spin correlations to have attained their ground
state values (see Figs.~\ref{fig:Saf0} and \ref{fig:SX01}). Figure
\ref{fig:Awbeta2} shows the density of states in the $f$-band as a
function of the interband hybridization $V$, at two values of on-site
repulsion $U=4$ (upper panels) and $U=8$ (lower panels).  Results for
four values of the hopping parameter in the correlated band,
$t_{f}\in\{-0.9,-0.2,+0.2,+0.9\}$ are given.  In green we plot the
edges of the non-interacting bands, defined by ${\rm Max}\left[
  E^0_-({\bf k})\right]$ and ${\rm Min}\left[E^0_+({\bf k})\right]$,
which delimit the band gap at $U=0$.

For $t_f=\pm 0.9$ and $U=4$ the DOS has a peak at small $V$ which
extends all the way down to $V=0$. The peak is absent at $U=8$ where a
fully formed Mott gap appears as $V$ approaches 0. Both points
indicate that $\beta=2$ is too high temperature for moment formation
and Mott-insulating behavior to develop at $U=4$.  On the other hand,
at $U=4$ and $t_f=\pm 0.2$, moments have formed and the $\omega=0$
feature that develops at finite $V$ is a Kondo resonance.  The
situation at $U=8$ differs because, as already remarked, the $V=0$
system is a Mott insulator (in the $f$-band) even at
$t_f=\pm0.9$. Hence we see clear signs of broad Kondo resonances at
intermediate $V$ for all four values of $t_f$.

\begin{figure*}
\centerline{\includegraphics[width=18cm,angle=0]{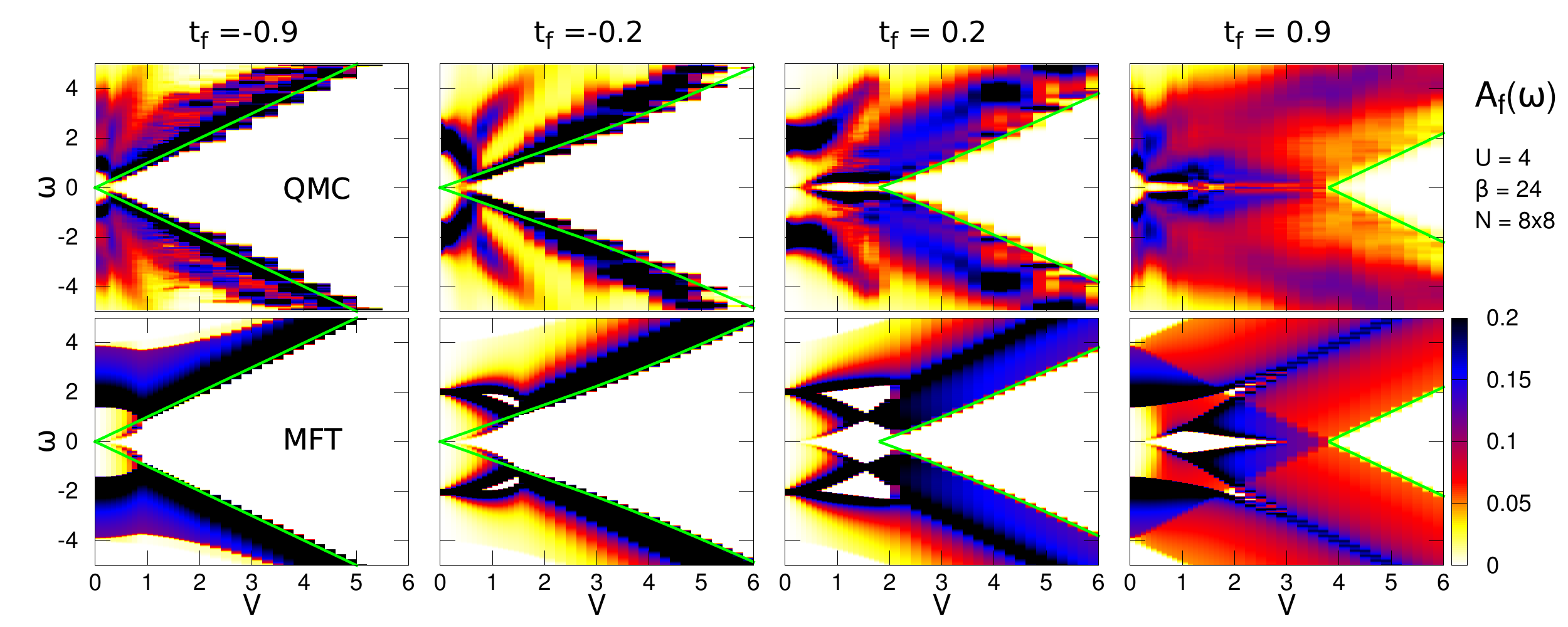}}
\caption{ The upper panels are the same as Fig.~\ref{fig:Awbeta2}
  except at much lower temperature $\beta=24$.  The smearing of the
  band gaps is greatly reduced.  In addition the Kondo resonance has
  been split by a gap associated with the formation of AF order.  The
  insulating character of the $t_f<0$ case for all $V$ is more
  apparent at this lower temperature.  We show only $U=4$, with QMC in
  the top panels and MFT in the bottom ones.  The green lines delimit
  the non-interacting band gap.  }
\label{fig:Awbeta24} 
\end{figure*}

While in the case of negative $t_{f}$ the non-interacting limit is
always gapped, in the case of positive $t_{f}$ there is a critical $V$
for insulating behavior, leading to a larger region of the phase space
in which low energy excitations allow Kondo physics to occur down to
low temperatures.  Alternatively put, the gapped single particle
density of states characteristic of negative $t_f$ freezes the
electrons and prevents screening of the $f$ moments.  This causes a
marked difference between the spectrum at positive and negative values
of $t_f$ at low temperature. Consider the cases of $t_f=\pm 0.2$ and
$\beta=24$ reported in Fig.~\ref{fig:Awbeta24}.  One immediately sees
that the sharp feature present at higher $T$ has survived for
$t_f=0.2$, albeit split as the result of the particular filling
considered and the onset of magnetic correlations. In contrast,
$t_f=-0.2$ shows no signature of the Kondo resonance that appears at
higher $T$.

It is also instructive to compare the QMC results with mean field
theory (Fig.~\ref{fig:Awbeta24}).  We observe that at the low
temperature presently considered, most of the features of the spectrum
are reproduced by the mean field solution.  At $t_f=0.2$ and $V<2$,
however, the mean field gap has a more pronounced dependence on $V$
and it is visibly larger. We attribute both differences to the fact
that DQMC describes a Kondo insulator, with a gap forming on top of
the Kondo resonance, while mean-field theory, by its very nature,
describes a Slater insulator and has no possibility to access the
physics of Kondo screening.

The particular way of representing the data makes also quite clear the
value of $V$ where one can identify the onset of band-insulating
behavior: a main feature visible in each of the panels is the
transition to a regime where the gap opens linearly as $V$ increases,
and whose width is consistent with the non-interacting expressions.
We found that for $t_f>0$, the onset of band insulating behavior
happens at roughly the same value of $V$ in both mean-field theory and
QMC. For $t_f<0$, there is a closing of the gap as the value of $V$ is
initially increased, so that one can identify the onset of
band-insulating behavior when this trend is reversed and the gap opens
up as $V$ further increases.  We found, especially clearly at
$t_f=-0.9$, that the value of $V$ at which such Mott-to-band insulator
``transition'' occurs is severely over-estimated by mean-field theory.

\section{Conclusions}

In this paper we have used Mean Field Theory and Determinant Quantum
Monte Carlo to explore systematically the effect of $f$-orbital
bandwidth in a two band Hubbard Hamiltonian. These data expand upon
the more well-studied cases when $t_f=0$ (the Periodic Anderson model)
and $t_f=t_d$ (the Hubbard bilayer).  In addition, we have obtained
data for $t_f$ of opposite sign to $t_d$.  Our work quantifies the
role that shifts in the overlap of localized orbitals on neighboring
atoms might play in the mechanism whereby pressure destroys AF
correlations in heavy fermion and transition metal oxide materials.

One key conclusion of our work is that magnetic correlations are
maximized for small positive $t_{f}$, in correspondence to where the
non-interacting system becomes metallic. 
We found that nearest-neighbor correlations show very peculiar behavior and are,
in fact, suppressed in the optimal regime for observing long-range order. 
We interpret this surprising behavior as a signature that the 
effective spin-spin interaction at the metal-insulator line is weaker than
in neighboring regions, but has a longer
range than for any other value of $t_f$.

Finally, the study of the single-particle spectral function reveals
that tuning the $f$-bandwidth deeply modifies the low-energy 
spectrum of this model. In contradiction to what was found in a
previous study \cite{JPSJ.69.1777}, the Coulomb interaction does
not simply lead to the renormalization of the non-interacting properties.
While certain of the qualitative features of the spectra can be well understood from
the mean-field and non-interacting band structure, other features
suggest more subtle correlation physics. In particular, at fairly high
temperature a Kondo resonance can develop at both positive and
negative $t_f$ values, while at low temperature only positive $t_f$
shows the presence of a split Kondo resonance.

\section{Acknowledgements}

S.C. acknowledge supports from DOE (DE-SC0008627).  This work was also
supported by the CNRS-UC Davis EPOCAL LIA joint research grant and by
the NNSA under Project \#201223433.


\begin{thebibliography}{100}

\bibitem{rasetti91}
{\it The Hubbard Model-  Recent Results}, M. Rasetti, World Scientific,
1991.

\bibitem{montorsi92}
``{\it The Hubbard Model},'' Arianna Montorsi (ed), World Scientific,
1992.

\bibitem{gebhard97}
{\it The Mott Metal-Insulator Transition, Models and Methods},
F.~Gebhard, Springer (1997).

\bibitem{fazekas99}
{\it Lecture Notes on Electron Correlation and Magnetism}, P.~Fazekas,
World Scientific (1999).

\bibitem{zaanen89}
J. Zaanen and O. Gunnarsson,
Phys. Rev. {\bf B40}, 7391 (1989).

\bibitem{machida89}
K. Machida, Physica C, {\bf 158}, 192 (1989)

\bibitem{kato90}
M. Kato, K. Machida, H. Nakanishi and M. Fujita, 
J. Phys. Soc. Jpn., {\bf 59}, 1047 (1990)

\bibitem{inui91}
M. Inui and P.B. Littlewood,
Phys. Rev. {\bf B44}, 4415 (1991).

\bibitem{yang91}
J. Yang and W.P. Su,
Phys. Rev. {\bf B44}, 6838 (1991).

\bibitem{scalapino94}
D.J. Scalapino, Does the Hubbard Model Have the Right Stuff?  in {\it
Proceedings of the International School of Physics} (July 1992), edited
by R. A. Broglia and J. R. Schrieffer (North-Holland, New York, 1994),
and references cited therein.

\bibitem{sugitani06}
I. Sugitani, Y. Okuda, H. Shishido, T. Yamada,
A. Thamizhavel, E. Yamamoto, T.D. Matsuda, Y.
Haga, T. Takeuchi, R. Settai and Y. Onuki,
 J. Phys. Soc. Jpn. {\bf 75} 043703 (2006).

\bibitem{stewart01}
G. R. Stewart, Rev. Mod. Phys.  {\bf 73}, 797 (2001).

\bibitem{georges96}
A. Georges, G. Kotliar, W. Krauth and M.J. Rozenberg
Rev. Mod. Phys. {\bf 68}, 13 (1996).

\bibitem{vollhardt93}
D. Vollhardt, in Correlated Electron Systems, edited by
    V. J. Emery (World Scientific, Singapore, 1993), p. 57;
    Th. Pruschke, M. Jarrell, J. Freericks, Adv. Phys. {\bf 44}, 187
    (1995). 

\bibitem{ulmke96}
M. Ulmke, R.T. Scalettar, A. Nazarenko, and E. Dagotto,
Phys. Rev. {\bf B54}, 16523 (1996).

\bibitem{held00}
K.~Held, C.~Huscroft, R.T.~Scalettar, and A.K.~McMahan,
Phys.~Rev.~Lett.~{\bf 85}, 373 (2000).

\bibitem{scalettar94}
R.T.~Scalettar, J.W.~Cannon, D.J.~Scalapino,
and R.L.~Sugar, Phys.~Rev.~{\bf B50}, 13419 (1994).

\bibitem{sandvik94}
A.W. Sandvik and D.J. Scalapino,
Phys.~Rev.~Lett.~{\bf 72}, 2777 (1994).

\bibitem{white92-98}
 S. R. White, Phys. Rev. Lett.  {\bf 69}, 2863 (1992); Phys.
Rev. {\bf B48}, 10345 (1993);
S. R. White and D. J. Scalapino, Phys. Rev. Lett.  {\bf 80}, 1272
(1998); Phys. Rev. Lett. {\bf 81}, 3227 (1998).

\bibitem{PhysRevB.76.155126}
A.~Toropova, C.~A. Marianetti, K.~Haule, and G.~Kotliar,
Phys. Rev. {\bf B76}, 155126 (2007).

\bibitem{PhysRevB.49.4432}
M.A. Continentino, G.M. Japiassu, and A. Troper,
Phys. Rev. {\bf B49}, 4432 (1994).

\bibitem{JPSJ.69.1777}
Y. Shimizu, O. Sakai, and A.C. Hewson,
J. Phys. Soc. of Japan, 
{\bf 69}, 1777 (2000).

\bibitem{JPSJ.74.2517}
O. Sakai, Y. Shimizu, and Y. Kaneta,
J. Phys. Soc. of Japan, {\bf 74}, 2517 (2005).

\bibitem{demedici05}
L. de' Medici, A. George, G. Kotliar, and S. Biermann,
Phys. Rev. Lett. {\bf 95}, 066402 (2005).

\bibitem{JPSJ.80.064710}
T. Yoshida, T. Ohashi, and N. Kawakami,
J. Phys. Soc. Japan, {\bf 80}, 064710 (2011).


\bibitem{liebsch04}
A. Liebsch, Phys. Rev. {\bf B70}, 165103 (2004).

\bibitem{liebsch05}
A.  Liebsch, Phys. Rev. Lett. {\bf 95}, 116402 (2005).

\bibitem{arita05}
R. Arita and K. Held, Phys. Rev. {\bf B72}, 201102(R) (2005).

\bibitem{inaba06}
K. Inaba and A. Koga, Phys. Rev. {\bf B73}, 155106 (2006).

\bibitem{costi07}
T.A. Costi
and A. Liebsch, Phys. Rev. Lett. {\bf 99}, 236404 (2007).

\bibitem{koga04}
A. Koga, N. Kawakami, T.M. Rice, and M. Sigrist, Phys. Rev. Lett. {\bf
92}, 216402 (2004).

\bibitem{ferrero05}
M. Ferrero, F. Becca, M. Fabrizio, and M. Capone,
Phys. Rev. {\bf B72}, 205126 (2005).

\bibitem{demedici05b}
L. de'Medici, A.
Georges, and S. Biermann, Phys. Rev. {\bf B72}, 205124 (2005).

\bibitem{ruegg05}
A. Ruegg M. Indergand, S. Pilgram and M. Sigrist, Eur. Phys. J. {\bf
  B48}, 55 (2005).

\bibitem{inaba05}
K. Inaba, A. Koga, S. Suga, and N. Kawakami, Phys. Rev. {\bf
B72}, 085112 (2005).

\bibitem{knecht05}
C. Knecht, N. Blumer, and P.G.J. van
Dongen, Phys. Rev. {\bf B72}, 081103(R) (2005).

\bibitem{biermann05}
S. Biermann, L. de'Medici, and A.
Georges, Phys. Rev. Lett. {\bf 95}, 206401 (2005).


\bibitem{blankenbecler81}
R. Blankenbecler, D.J. Scalapino, and R.L. Sugar, Phys. Rev. {\bf D24},
2278 (1981).

\bibitem{white89}
S.R.~White, D.J.~Scalapino, R.L.~Sugar,
E.Y.~Loh, Jr., J.E.~Gubernatis, and R.T.~Scalettar,
Phys.~Rev.~{\bf B40}, 506 (1989).

\bibitem{trotter}
H.F. Trotter, Proc. Amer. Math. Soc. {\bf 10}, 545 (1959); M.
Suzuki,
Prog. Theor. Phys. {\bf 56}, 1454 (1976);
R.M. Fye, Phys. Rev.  {\bf
B33}, 6271 (1986); and
R.M. Fye and R.T. Scalettar, Phys. Rev. {\bf B36},
3833 (1987).

\bibitem{varney09}
C.N. Varney, C.R.~Lee, Z.J. Bai, S. Chiesa, M. Jarrell, and R. T.
Scalettar, Phys. Rev. {\bf B80}, 075116 (2009).

\bibitem{gubernatis91}
J.E. Gubernatis, M. Jarrell, R.N. Silver, and D.S. Sivia,
Phys. Rev. {\bf B44}, 6011 (1991).

\bibitem{beach04}
K.S.D.~Beach, arXiv preprint cond-mat/0403055 (2004).

\bibitem{vekic95}
M.~Vekic, J.W.~Cannon, D.J.~Scalapino, R.T.~Scalettar,
and R.L.~Sugar, Phys.~Rev.~Lett.~{\bf 74}, 2367 (1995).


\end{thebibliography}
\end{document}